\documentclass[11pt]{JHEP3}

\def\jt{\tilde{j}}
\def\mt{\tilde{m}}
\def\Nt{\tilde{N}}
\def\ni{\noindent}

\def\TL{\hfil$\displaystyle{##}$}
\def\TR{$\displaystyle{{}##}$\hfil}

\def\TT{\hbox{##}}

\def\lbldef#1#2{\expandafter\gdef\csname #1\endcsname {#2}}
\def\eqn#1#2{\lbldef{#1}{(\ref{#1})}%
\begin{equation} #2 \label{#1} \end{equation}}
\def\eqalign#1{\vcenter{\openup1\jot
    \halign{\strut\span\TL & \span\TR\cr #1 \cr
   }}}

\def\href#1#2{#2}

\newcommand{\beq}{\begin{equation}}
\newcommand{\eeq}{\end{equation}}
\newcommand{\ber}{\begin{eqnarray}}
\newcommand{\eer}{\end{eqnarray}}

\newcommand{\beqar}{\begin{eqnarray}}

\newcommand{\cH}{{\cal H}}

\newcommand{\cG}{{\cal G}}

\newcommand{\cN}{{\cal N}}

\newcommand{\cF}{{\cal F}}

\newcommand{\cC}{{\cal C}}

\def\cD{{\cal D}}
\def\cCo{{\cal C}}

\newcommand{\eeqar}{\end{eqnarray}}

\title{Unitarity of supersymmetric $SL(2,R)/U(1)$ and
no-ghost theorem for fermionic strings in $AdS_3 \times \cN$}
\author{Ari Pakman \\ Racah Institute of Physics, The Hebrew University, Jerusalem 91904, Israel \\
E-mail: \email{pakman@phys.huji.ac.il}}

\abstract{The unitarity of the NS  supersymmetric coset $SL(2,R)/U(1)$ is studied
for the discrete representations. The results are applied to the proof of the no-ghost theorem for
fermionic strings in $AdS_3 \times \cN$ in the NS sector.
A no-ghost theorem is proved for states in flowed discrete representations.}

\begin{document}

\baselineskip=15.5pt

\pagestyle{plain}

\setcounter{page}{1}

\section{Introduction}
The purpose of this work is to study the unitarity of the
supersymmetric Euclidean $SL(2,R)/U(1)$ coset,
which is the simplest example of a nontrivial noncompact Kazama-Suzuki
model \cite{KS}.
The coset is with respect to the timelike $U(1)$ current of
$SL(2,R)$.
Our result  is a supersymmetric generalization
of similar results for the  bosonic coset \cite{Dixon}, and
is relevant for several models.
We will use it here to fill a gap in the proof of the no-ghost
theorem \cite{Evans} for fermionic strings in $AdS_3 \times N$.
We will also extend the proof to fermionic string excitations
belonging to the flowed sectors introduced in \cite{Maldacena}.

The bosonic coset $SL(2,R)/U(1)$ is a non-linear sigma model
whose target space geometry to first order in $1/k$ is that of a 2D black hole \cite{Witten,DVV}.
In its supersymmetric version,  the 2D black hole metric
has been shown to describe the {\it exact} conformal background up to four loops \cite{Jack},
and has been argued to be exact to all orders \cite{Tseytlin}.
Generalizing similar results for the bosonic case,
the supersymmetric coset has been argued
to be equivalent to certain matrix models \cite{WittenMatrix} and to $N=2$ Liouville theory \cite{Giveon, Hori},
and is also relevant in relation to Little String Theories \cite{Giveon}.

The unitarity challenges presented by $AdS_3$ are very old
\cite{Balog:1989jb,
Petropoulos:1990fc,
Mohammedi:1990dp,
Bars:1991rb,
Hwang:1991aq,
Hwang:1998tr}.
A  proof of the no-ghost theorem for bosonic strings
on $AdS_3$ was given in \cite{Evans}, based on \cite{Hwang}.
A different approach to unitarity for strings in $AdS_3$ has been advocated in \cite{Bars}.

The relation between the $SL(2,R)/U(1)$ coset and string unitarity in $AdS_3 \times \cN$
comes from the fact that the proof of the no-ghost theorem \cite{Hwang,Evans}
shows that physical states built upon $SL(2,R)$
representations lie in the coset, modulo spurious states\footnote{As shown in Section 4, this statement does not hold
for states with $J^3_0$ eigenvalue 0, which should be treated separately}.
This was shown in \cite{Evans} both for bosonic and  fermionic strings.
So string unitarity necessitates coset unitarity.
The latter was proved in \cite{Dixon} for the bosonic coset, and we
will prove it for the supersymmetric case in the discrete representations.
It is worth noticing that both string and coset unitarity
rely on truncating the spin $j$ of the discrete representations. This
restriction of $j$ is rather ubiquitous, and we discuss it in Section 2.

The plan of this work is as follows. In Section 2 we review the basics
of string quantization in $AdS_3$. In Section 3 we prove the unitarity of the coset. In Section 4 we review
the no-ghost theorem for unflowed string states in the discrete series \cite{Evans} and extend it
to the flowed sectors. Section 5 contains the conclusions.

In this work we will only deal with the Neveu-Schwarz sector. The Ramond sector can be treated along the same lines.

\section{Fermionic strings in $AdS_3$ and spectral flow}

The worldsheet theory is composed by a supersymmetric $SL(2,R)$ WZW model at level $k$ with central charge
\eqn{cargasl}{
c_{SL(2,R)}=3 + \frac{6}{k} + \frac32
}
and a supersymmetric unitary theory $\cN$ with central charge $c_{\cN}=15-c_{SL(2,R)}$.

The current algebra of supersymmetric $\widehat{SL}(2,R)$ at level k is generated by six currents  $J^{3,\pm} , \psi^{3,\pm}$,
whose modes satisfy
\eqn{comm}{
\eqalign{
[J^3_n, J^3_m ]    &= - {k \over 2} n \delta_{n+m,0} \cr
          [J^3_n, J^\pm_m ]  &= \pm J^\pm_{n+m} \cr
          [J^+_n , J^-_m ]   &= -2J^3_{n+m} + kn\delta_{n+m,0}  \cr
      [J^3_n, \psi^\pm_m]&=\pm \psi^\pm_{n+m} \cr
      [J^\pm_n,\psi^\mp_m]&=\mp 2\psi^3_{n+m} \cr
      [J^\pm_n, \psi^3_m]&=\mp \psi^\pm_{n+m} \cr
      \{ \psi^3_n,\psi^3_m \}&= -{k \over 2} \delta_{n+m,0} \cr
      \{ \psi^+_n,\psi^-_m \}&= k \delta_{n+m,0},
}}
with all other (anti)commutators vanishing, and the same for the antiholomorphic sector.
The modding of $J^a_n$ is integer and that of $\psi^a_n$ half-integer (Neveu-Schwarz) or integer (Ramond).
As usual, purely bosonic currents can be defined by
\eqn{bosonicJ}{
\eqalign{ j^a &= J^a - \hat{J}^a   \cr
      \hat{J}^a &= -{i \over k} f^a_{~bc} \psi^b \psi^c
}}
The $j^a$ form a $\widehat{SL}(2,R)$ bosonic algebra at level $k+2$.
The $\hat{J}^a$ and the $\psi^a$ form a supersymmetric $\widehat{SL}(2,R)$ algebra
at level $-2$ which commutes with $j^a$. The spectrum
of the theory is the direct product of the Hilbert spaces of both theories.

The Sugawara stress tensor and the supercurrent are \eqn{tensorT}{
\eqalign{ T &={1 \over 2k}[j^+j^- +j^-j^+] - {1 \over k}j^3j^3 -{1
\over 2k}[\psi^+ \partial \psi^- + \psi^- \partial \psi^+] +{1
\over k}\psi^3 \partial \psi^3 \cr G &={1 \over k} [\psi^+j^- +
\psi^-j^+] -{2 \over k}\psi^3j^3 - {2 \over k^2}
\psi^+\psi^-\psi^3 }} \noindent In particular the zero mode of $T$
is \eqn{modocero}{ \eqalign{ L_0 & = {1 \over
k}[{1\over2}(j^+_0j^-_0 +j^-_0j^+_0) - j^3_0j^3_0 +
\sum_{m=1}^\infty j^+_{-m}j^-_m +j^-_{-m}j^+_m - 2j^3_{-m}j^3_m]
\cr & + {1 \over k}\sum_{m={1\over2}}^\infty m(\psi^+_{-m}
\psi^-_m + \psi^-_{-m} \psi^+_m -2 \psi^3_{-m} \psi^3_m) }} \ni
The highest weight representations of \comm\ are built as a direct
product of representations of $j^a$ and $\psi^a$. For the $\psi^a$
currents, we have the usual representations for free fermions, for
both Neveu-Schwarz and Ramond sectors. For the NS sector, in which
we will be interested, we have a vacuum $|0 \rangle$ annihilated
by $\psi^a_{n>0}$ upon which the $\psi^a_{n<0}$ states act.

For $j^a$, we start from unitary representations $|j;t \rangle$ of
the $SL(2,R)$ Lie algebra  $j^{3,\pm}_0$. The representations are
characterized by the eigenvalues $-j(j-1)$ of the Casimir operator
$\frac12 (j^+_0 \! j^-_0  \! + \! j^-_0 \! j^+_0 ) \! - \!
(j^3_0)^2$. The states within each representation are labelled by
the eigenvalue $t$ of $j^3_0$. The states $|j;t \rangle$ are
annihilated by $j^{3,\pm}_{n>0}$, and the Fock space of states is
built by acting on them with $j^{3,\pm}_{n<0}$.

The unitary representations of the $SL(2,R)$ Lie algebra $j^{3,\pm}_0$
appearing in the spectrum of strings moving in an $AdS_3 \times \cN$ background  \cite{Maldacena} are:

\begin{enumerate}
\item {\bf Highest weight discrete representations}
\[ \cD_j^+ = \left\{ |j;t \rangle : t=j,j+1, j+2, \cdots \right\} \]
where $|j;j \rangle$ is annihilated by $j_0^-$ and $j$ is a  real number such that
$1/2 < j < (k+1)/2$.

\item {\bf Lowest weight discrete representations}
\[ \cD_j^- = \left\{ |j;t \rangle : t=-j,-j-1, -j-2, \cdots \right\} \]
where $|j;-j \rangle$ is annihilated by $j_0^+$ and $j$ is a  real number such that
$1/2 < j < (k+1)/2$.

\item {\bf Continuous representations}
\[ \cCo_j = \left\{ |j,\alpha;t \rangle : t= \alpha, \alpha \pm 1,\alpha \pm 2, \cdots \right\} \]
where $0 \leq \alpha < 1$ and $j = 1/2 + \textrm{i} z $, where $z$
is a real number.
\end{enumerate}

\ni The  bounds for  $j$ appearing in $\cD_j^{\pm}$ can be
understood in terms of consistency conditions for the primary
states. The lower bound, $ 1/2 < j$,  is necessary for the
normalizability of the primary states when their norm is
interpreted as the ${\cal L}^2$ inner product of functions in the
$SL(2,R)$ group manifold \cite{DVV}. As for the upper bound, $ j <
(k+1)/2$,  it was noted in \cite{Giveon} that it is necessary for
the unitarity of the primary states, when their norm is
interpreted as the two-point function of the vertex operators
creating them from the vacuum.

Moreover, adopting either the upper or the lower bound for $j$, the other one
appears when imposing the $w=\pm 1$ spectral flow (see below) to be a symmetry of the spectrum.
Finally, the compelling evidence for the correctness of these bounds on $j$ comes from
the fact that only this range of $j$ appears in the spectrum
of the thermal partition function of the model, computed by
path integral techniques in \cite{Maldacena}.

Now, given any integer $w$, the algebra \comm\ is preserved by the spectral flow
$J_n^{3,\pm} \rightarrow \tilde{J}_n^{3,\pm}$, $\psi_n^{3,\pm} \rightarrow \tilde{\psi}_n^{3,\pm}$
defined by
\eqn{flow}{
\eqalign
{
  \tilde{J}_n^3 & = J_n^3 - {k \over 2} w \delta_{n,0} \cr
  \tilde{J}_n^\pm & = J_{n \pm w}^\pm \cr
  \tilde{\psi}_n^3 & = \psi_n^3 \cr
  \tilde{\psi}_n^\pm & = \psi_{n \pm w}^\pm
}} For $w=\pm 1$, the symmetry \flow\ maps the $\cD^{\mp}_j$
representation into $\cD^{\pm}_{k/2-j}$. But for generic~$w$, it
was shown in \cite{Maldacena}  that this symmetry maps regular
$SL(2,R)$ representations to new representations which must be
included in a consistent quantization of strings in $AdS_3 \times
\cN$.

Note that the spectral flow was also performed on the fermions for consistency of the algebra.
The implications of this simultaneous flow for the boundary CFT theory have
been discussed in \cite{AGS}.
But the new sectors can be obtained by performing the flow only on the purely bosonic sector $j^a$.
For the free fermions, it is easily seen to be a rearrangement
of the spectrum. In fact, for $w>0$ a flowed NS vacuum can be defined by
$$|\tilde{0}\rangle = \psi^-_{-{1 \over 2}} \psi^-_{-{3 \over 2}} \cdots \psi^-_{-|w|+ {1 \over 2}}|0\rangle \, , $$
while for $w<0$ the $\psi^-_n$ should be replaced by $\psi^+_n$. In both cases the new vacuum satisfies
$$\tilde{\psi}^{\pm,3}_n |\tilde{0} \rangle =0, ~~~~~ n>0,$$

\ni The flow \flow\ maps the modes of $T$ and $G$ to
\eqn{tgflowed}{ \eqalign{ \tilde{L}_n & =L_n + wJ^3_n -
{\frac{k}4}w^2\delta_{n,0} \cr \tilde{G}_r & =G_r + w\psi^3_r }}
Physical states will satisfy, in the $-1$ picture \cite{FMS},
\eqn{phys}{ (L_{n \geq 0}-{1 \over 2}\delta _{n,0}) |x\rangle =
G_{n>0}|x\rangle=0 } The way to include the flowed representations
is to impose the physicality conditions \phys, with  {\em
unflowed} $L_n$ and  $G_n$, on $\tilde{j}^a$ descendants built
upon $|\tilde{j},\tilde{t} \rangle$ representations.

For generic $w$, the mass shell condition can be
expressed as
\eqn{shellmass}{
-{\jt ( \jt -1) \over k} -wm + {kw^2 \over 4}+ N + h - {1 \over 2}=0
}
where  $h$ a highest weight
of the inner $\cN$ theory and  the level $N$ is a half-integer number.

Since the spin in the continuous representation of $SL(2,R)$ is
$j= \frac12 + \textrm{i} z$ with $z$ real, the mass shell
condition for unflowed strings is \eqn{masslong}{ {{\frac14 + z^2}
\over k} +N+h - {1 \over 2}=0 } which can only be satisfied for
$N=0$. So string states in this sector will always be unitary
because the zero mode representation is. This does not hold for
flowed states in $\cC_j$, which should be treated separately (see
Section 5).

\section{Unitarity of  supersymmetric $SL(2,R)/U(1)$}

The states in the $SL(2,R)$ module
can in general have negative norm because the currents $J^3, \psi^3$ are timelike.
The states of the coset
are states $|x\rangle$ such that
$J^3_{n>0}|x\rangle=\psi^3_{n>0}|x\rangle=0$
In this section we will show that the coset states built upon
the $\cD_j^{\pm}$  representations of the zero modes, have positive norm provided $k>2$ and $j<{k \over 2} + 1$.
The proof is a generalization to the supersymmetric case of
\cite{Dixon} for the bosonic coset.

The no-ghost theorem requires unitarity of states annihilated by
$J^3_{n>0}$ and $\psi^3_{n>0}$. But if we were interested instead
in proving the unitarity of states annihilated by $j^3_{n>0}$ and
$\psi^3_{n>0}$ the result would follow immediately from that of
the bosonic case, due to the direct product structure. A simple
example of a state annihilated by $J^3_{n>0}$ and $\psi^3_{n>0}$
but not by  $j^3_{n>0}$ is given by \eqn{ejemplo}{
|x\rangle=J^3_{-1}|j,t\rangle +{1 \over
2}\psi^+_{-1/2}\psi^-_{-1/2}|j,t\rangle }

Moreover, only the condition imposed with $J^3_{n>0}$ is
consistent with supersymmetry in the sense that if we require of
our state to be simultaneously a primary of the supercurrent $G$,
no further restrictions arise. If the conditions are imposed with
$j^3_{n>0}$, it follows that the state should be also annihilated
by $\hat{J}^3_{n \geq 0}$.

Before addressing the
unitarity issue itself, we should first discuss the quantum numbers
characterizing a state belonging to the supersymmetric $SL(2,R)/U(1)$ coset.
This is the major subtlety in generalizing the result of \cite{Dixon}.

Consider the decompositions
\eqn{decom}{
\eqalign{
J^3_0 & = j^3_0 + \hat{J}^3_0 \cr
L_0 & = L_0^b + L_0^f
}}
where $L_0^b$ and $L_0^f$ are the zero modes of
\eqn{tsepdef}{
\eqalign{
T^b &= {1 \over 2k}[j^+j^- +j^-j^+] - {1 \over k}j^3j^3 \cr
T^f &=-{1 \over 2k}[\psi^+ \partial \psi^- + \psi^- \partial \psi^+]
+{1 \over k}\psi^3 \partial \psi^3 \,\,\,,
}}
respectively, with $T=T^b + T^f$.

Due to the direct product structure, a basis can be chosen for the $SL(2,R)$ module
made up of states diagonal in  $j^3_0, \hat{J}^3_0, L_0^b$ and $L_0^f$.
The eigenvalues of $j^3_0$ and $\hat{J}^3_0$ will be called $t$ and $s$, and the
levels of $L_0^b$ and $L_0^f$ will be denoted by    $N^b$ and $N^f$.
Calling $m$ and $N$ the
eigenvalue of $J^3_0$ and the level of $L_0$,  \decom\ implies $m=t+s$ and $N=N^b + N^f$.

A state belonging to the coset can be assumed to have definite $N,
t$ and $s$ quantum numbers, because the commutators
\eqn{diago}{\eqalign{ [L_0,J^3_n] & =-nJ^3_n \cr [L_0,\psi^3_n] &
=-n\psi^3_n \cr [j^3_0,J^3_n]& =[j^3_0,\psi^3_n] =0 \cr
[\hat{J}^3_0,J^3_n]& =[\hat{J}^3_0,\psi^3_n] =0 }} imply that
$J^3_{n>0}$ and $\psi^3_{n>0}$ annihilate states with different
$N,t$ or $s$ separately.

On the other hand, a state in the coset will not in general have definite $N^b$ and $N^f$ quantum numbers,
because the nondiagonal commutators
\eqn{nodiago}{\eqalign{
[L_0^b,J^3_n]=-nj^3_n \cr
[L_0^f,J^3_n]=-n\hat{J}^3_n
}}
show that action of $J^3_n$ and $\psi^3_n$ will generally mix
states with different $N^b$ and $N^f$.
For example, rewriting the level $N=1$ coset state \ejemplo\ as
\eqn{ejemplor}{
|x\rangle=j^3_{-1}|j,t\rangle +\left({1 \over k}+{1 \over 2}\right)\psi^+_{-1/2}\psi^-_{-1/2}|j,t\rangle
}
it is clear that it has $s=0$ and $t=m$, but $N^b$ and $N^f$ are different for the two terms.

Let's call $N^b_{max}$ to the maximum value of $N^b$ appearing in the
terms of the expansion of a coset state. In  the same term will appear $N^f_{min}=N-N^b_{max}$ as
the minimum value of $N^f$. It is clear that the structure of the algebra implies that $2N^f_{min} \geq s^2$ and $t \geq j-N^b_{max}$.

\medskip
\medskip
Let's see now that the norm of $|x\rangle$ is positive. Consider
\eqn{level}{
\langle x| L_0 |x\rangle= \left(-{j(j-1) \over k} + N \right)\langle x|x\rangle
}
We will use now the explicit form \modocero\ of $L_0$, but we write
the terms that come from $j^3j^3$ in \tensorT\ as coming from
$$j^3j^3=J^3J^3 - \hat{J^3}J^3 - J^3\hat{J^3} + \hat{J^3}\hat{J^3},$$
\noindent
whose zero mode is
\eqn{zerojj}{
(J_0^3)^2  -2J_0^3\hat{J}^3_0 + 2\sum^\infty_{n=1}(J^3_{-n}J^3_{n} -\hat{J}^3_{-n}J^3_{n}-J^3_{-n}\hat{J}^3_{n})
 + {2 \over k}\sum_{q={1\over2}}^\infty q(\psi^+_{-q} \psi^-_q + \psi^-_{-q} \psi^+_q) \,\,,
}
where we have used
$$\hat{J}^3\hat{J}^3= {1 \over k^2}{:\psi^+\psi^-:}{:\psi^+\psi^-:}=
 - {1\over k} \psi^+ \partial \psi^- - {1\over k} \psi^- \partial \psi^+ \,.$$

\noindent
The resulting expression for $L_0$ is
\begin{eqnarray}
\label{elecero}
L_0 &=& {1 \over k}[{1\over2}(j^+_0j^-_0 +j^-_0j^+_0) +
\sum_{m=1}^\infty j^+_{-m}j^-_m +j^-_{-m}j^+_m] \nonumber
\\
&+& {1 \over k}[(J_0^3)^2  -2J_0^3\hat{J}^3_0 + 2\sum^\infty_{n=1}(J^3_{-n}J^3_{n} -\hat{J}^3_{-n}J^3_{n}-J^3_{-n}\hat{J}^3_{n})] \nonumber
\\
&+& {1 \over k}\sum_{m={1\over2}}^\infty m[(1-2/k)(\psi^+_{-m} \psi^-_m + \psi^-_{-m} \psi^+_m)
-2 \psi^3_{-m} \psi^3_m]
\end{eqnarray}

\ni
Inserting this expression in \level\ and
annihilating the modes $J^3_{n>0},\psi^3_{n>0} \, (J^3_{n<0},\psi^3_{n<0})$ against
$|x\rangle \,(\langle x|)$,  \level ~can be rearranged into
\eqn{norma}{
\langle x|x\rangle=
{{\langle x|[ \sum_{p=0}^\infty j^+_{-p}j^-_p + \sum_{p=1}^\infty j^-_{-p}j^+_p
+ \left( 1-{2 \over k} \right) \sum_{q={1\over2}}^\infty q(\psi^+_{-q} \psi^-_q + \psi^-_{-q} \psi^+_q)]|x\rangle}
\over {t(t-1)-j(j-1)+kN-s^2}}
}

We will see that $\langle x|x\rangle$ is positive by showing that both the numerator and the denominator of \norma\ are
positive.

Let's consider first the denominator. When $t \geq j$ it can be written as
\eqn{denmayor}{
t(t-1)-j(j-1) + kN^b_{max} + (2N^f_{min}-s^2) + (k-2)N^f_{min}
}
which is positive for $k > 2$. When $t<j$, we rewrite it as
\eqn{denmenor}{
k(N^b_{max}+t-j) + (j-t)(j-t-1)+2(k/2+1-j)(j-t) + (2N^f_{min}-s^2) + (k-2)N^f_{min}
}
which is also positive for $k>2$ and $j < {k \over 2}+1$. So the denominator is always positive.

\medskip

\medskip
Regarding the numerator, we proceed by induction on $N$ and $m$.
We assume that unitarity holds for states in
the coset module with level lower than $N$, or $J_0^3$ eigenvalue lower than $m$.

Note that the numerator is the sum of norms of states
of the form $j^\pm_{p>0}|x\rangle, \psi^\pm_{q>0}|x\rangle$ all of which
have level lower than $N$, and $j^-_0|x\rangle$ which has
level $N$ but lower $J^3_0$ eigenvalue.
But the norm of these states is not guaranteed to be positive by the induction hypothesis,
since, although they are still annihilated by $\psi^3_{r>0}$, they are not necessarily annihilated by $J^3_{n>0}$.

Now, let {\bf P} be the projection operator on highest weight
states of $J^3$. Since every state of the theory can be obtained
by applying lowering operators $J^3_{n<0}$ to highest weight
states, we have the completeness relation

\eqn{complet}{
\eqalign{
1={\bf P} + & \sum_{n>0}\left( -{2\over kn} \right) J^3_{-n} {\bf P} J^3_{n} \cr
& + {1 \over 2!}\sum_{n_1,n_2>0} \left( -{2\over kn_1}\right) \left( -{2\over kn_2}\right) J^3_{-n_1}J^3_{-n_2} {\bf P}
J^3_{n_2}J^3_{n_1}+ ...
}}
\ni
Inserting this expression in norms like $\langle x|j^+_{-p}j^-_p|x\rangle$ and
$\langle x|\psi^+_{-p}\psi^-_p|x\rangle$ infinite
series are obtained, typical terms being

\eqn{tipico}{
{
{1 \over d!} \left(-{2\over k}\right)^d {1 \over {n_1...n_d}} \langle x|j^+_{-p} J^3_{-n_1}...J^3_{-n_d} {\bf P} J^3_{n_d}...J^3_{n_1}
j^-_p|x\rangle}
}

\eqn{tipico2}{
{
{1 \over d!} \left(-{2\over k}\right)^d {1 \over {n_1...n_d}} \langle x|\psi^+_{-q} J^3_{-n_1}...J^3_{-n_d} {\bf P} J^3_{n_d}...J^3_{n_1}
\psi^-_q|x\rangle}
}
\ni
Commuting the $J^3_n$ and annihilating them against $|x\rangle$ and $\langle x|$, the above expressions
can be brought to
\eqn{produce1}{
{
{1 \over d!} \left(-{2\over k}\right)^d {1 \over {n_1...n_d}} \langle x|j^+_{-(p+n_1+...+n_d)} {\bf P}
j^-_{(p+n_1+...+n_d)}|x\rangle}
}
\eqn{produce2}{
{
{1 \over d!} \left(-{2\over k}\right)^d {1 \over {n_1...n_d}} \langle x|\psi^+_{-(q+n_1+...+n_d)} {\bf P}
\psi^-_{(q+n_1+...+n_d)}|x\rangle}
}
Note that the $\langle x| \cdots |x\rangle$ factors are positive by the induction hypothesis. Using this
method, the numerator of \norma\ can be put into the form
\eqn{sumado}{
\langle x|[ \sum_{p=0}^\infty F_p(y)j^+_{-p}{\bf P}j^-_p + \sum_{p=1}^\infty F_{p-1}(y)j^-_{-p}{\bf P}j^+_p
+ (1- y) \sum_{q={1\over2}}^\infty H_q(y)(\psi^+_{-q}{\bf P} \psi^-_q + \psi^-_{-q}{\bf P} \psi^+_q)]|x\rangle
}

\noindent
where $y = {2 \over k}$ and
\eqn{efeyge}{
\eqalign{
F_p(y) & = \sum^p_{r=0}f_r(y), \cr
H_q(y) & = \sum^{q-1/2}_{r=0}(q-r)f_r(y)
}}

\noindent
where $f_0(y)=1$ and
\eqn{efecomosuma}{
f_{r \geq 1}(y)=\sum_{d=1}^\infty {(-y)^d \over d!} \sum_{n_i>0}{1 \over {n_1...n_d}} \delta_{n_1+n_2+...n_d,r}
}

\noindent
It is clear that \sumado\ will be positive if $F_p(y)$ and $H_q(y)$ are positive and $y<1$.
In order to express $F_p(y)$ and $H_q(y)$ in a closed form, we first consider, following \cite{Dixon},
the generating function for $f_r(y)$

\eqn{generatriz}{
\eqalign{
f(y,z)= & \sum_{r=0}^\infty f_r(y)z^r \cr
= & 1 + \sum_{d=1}^\infty {(-y)^d \over d!} \prod_{i=1}^d \left( \sum_{n_i=1}^\infty {z^{n_i} \over n_i}\right) \cr
= & \sum_{d=0}^\infty {(-y)^d \over d!} \left[ -\ln (1-z) \right]^d \cr
= & (1-z)^y
}}

\noindent
Now $f_r(y)$ can be obtained by derivation to find

\eqn{efechico}{
\eqalign{
f_1(y) & =-y, \cr
f_{r>1}(y) & = - {y \over r}(1-y)(1- {y \over 2})...(1-{y \over r-1})
}}

\noindent
and from \efeyge\ we obtain $F_0(y) =1$ and

\eqn{efegrande}{
F_{p \geq 1}(y) =(1-y)(1- {y \over 2})...(1-{y \over p})
}

\noindent
which is positive for $y<1$. Regarding $H_q(y)$ we see from \efeyge\ and \efechico\ that
\eqn{ge}{
\eqalign{
H_{1 \over 2}(y) & = 1/2 \cr
H_{3 \over 2}(y) & = {3 \over 2}F_1(y) + y \cr
H_{q>{3 \over 2}}(y) &  = qF_{q-{1 \over 2}} + y + y(1-y)+...+y(1-y)(1- {y \over 2})...(1-{y \over q-{3 \over 2}})
}}

\noindent
so that $H_q(y)$ is also positive for $y<1$.
It follows then that \sumado ~is positive for $y<1$, that is, for $k>2$.

\section{No-ghost theorem for fermionic strings in the discrete series of the NS sector}

The no-ghost theorem guarantees that states satisfying the physicality conditions
\phys\ have positive norm.
For the fermionic string, it was shown in \cite{Evans} that physical states
belong to the supersymmetric $SL(2,R)/U(1)$ coset, modulo spurious states. To this statement
the result of Section 3 must be added. For the spectrally flowed representations of $\widehat{SL}(2,R)$
a proof was given in \cite{Maldacena} for the
bosonic case.

In Section 4.1, for the sake of completeness, we will review the proof of \cite{Evans} for
unflowed fermionic strings, and then
generalize it to the flowed case in Section 4.2. We
will work out the details of the $\cD_j^+$
representations, the $\cD_j^-$ case being similar.

\subsection{Unflowed states}
We define the subspace $\cal{F}$ of the Hilbert space as consisting of states $|f\rangle$
which are primaries of $J^3(z), \psi^3(z), T(z)$ and $G(z)$, that is,
\eqn{primaries}{
J^3_n|f\rangle= \psi^3_n|f\rangle=L_n|f\rangle=G_n|f\rangle=0,
~~~~n>0.
}

\noindent
Notice that by the result of Section 3, states in $\cal{F}$ have positive norm. The proof has three steps.

\medskip
\noindent
{\bf Step 1}: If $|N,m,\nu\rangle$ is an orthogonal basis of $\cal{F}$ at level $N$ and
$J^3_0$ eigenvalue $m$,
the states of
the form
\eqn{hilb}{
\eqalign{
 & \, G_{-1/2}^{\varepsilon_0} \cdots G_{-a-1/2}^{\varepsilon_a}\,
L_{-1}^{\lambda_1} \cdots L_{-d}^{\lambda_d} \cr
& ~~ (\psi^3_{-1/2})^{\delta_0} \cdots (\psi^3_{-b-1/2})^{\delta_b} \,
(J^3_{-1})^{\mu_1} \cdots (J^3_{-n})^{\mu_n} |N,m,\nu\rangle \,,
}}

\noindent
with $m$ fixed and $\nu$ varying, are linearly independent and
form a basis of the Hilbert space for states with level $M \geq N$
and $J^3_0$ eigenvalue $m$.

\bigskip
\noindent
{\bf Proof}:
Defining
\eqn{TT}{
\eqalign{
T^3 & = -{1 \over k} J^3J^3 - {1 \over k} \psi^3 \partial \psi^3     \cr
G^3 & = -{2 \over k} J^3 \psi^3,   }}

\noindent
the fields $J^3 , \psi^3, T^3$ and $G^3$ form a $c= {3 \over 2}$ supersymmetric timelike $U(1)$
algebra.
We can further define
\eqn{Tc}
{
\eqalign
{
L^c_n & =L_n-L_n^3   \cr
G^c_n & =G_n-G^3_n,
}}

\noindent
which, by construction, commute with $\psi^3_n$ and $J^3_n$.
Moreover,  $L^c_n$ and $G^c_n$ form a \^c=9 $(c= {27 \over 2})$
supersymmetric Virasoro algebra.
It is clear from \primaries , \TT \, and \Tc , that $|N,m,\nu\rangle$ is
also a primary of $L^c_n$, its
weight with respect to
${L^c}_0$ being,

\eqn{pesomax}{
h^c = - {j(j-1) \over k} + N + {m^2 \over k} + h\,,}

\noindent where $h$, is the highest weight of the internal unitary
$\cN$ CFT.

Using \Tc \, we can put the states \hilb \,
in one-to-one correspondence with states
\eqn{hilbc}{
\eqalign{
&(G_{-1/2}^c)^{\varepsilon_0} \cdots (G_{-a-1/2}^c)^{\varepsilon_a}\,
(L_{-1}^c)^{\lambda_1} \cdots (L_{-d}^c)^{\lambda_d}     \cr
& ~~ (\psi^3_{-1/2})^{\delta_0} \cdots (\psi^3_{-b-1/2})^{\delta_b} \,
(J^3_{-1})^{\mu_1} \cdots (J^3_{-n})^{\mu_n} |N,m,\nu\rangle \,,
}}
\ni
We want to show that for each $m$, the states \hilbc\ are a basis for the whole
Hilbert space. Linear independence of the states \hilbc\ is equivalent to the
non-singularity of the determinant of their inner products. Since states built upon different
$|N,m,\nu\rangle$'s or with different $(\psi^3_n, J^3_n)$
content are orthogonal, the determinant  factorizes into
Kac determinants of the $L^c_n,G^c_n$ supersymmetric Virasoro algebra
with highest weight $h^c$ and \^{c}=9.
These determinants in the NS sector are singular only for $h^c \leq 0$ \cite{Kac},
so we need $h^c > 0$.
Since the structure of the algebra implies that $m \geq j-N- 1/2$, rewriting \pesomax \, as

\eqn{pesomax2} {
 h^c = {(m-j)^2 \over k} + {2j \over k}(N + m + 1/2 - j) + {2 N \over k}({k \over 2} -j)+h,
}

\noindent
we see that it is strictly positive if
 $j < {k \over 2}$. Rewriting further \pesomax\ as
\eqn{pesomax3}{
h^c= {1 \over k}[(j- {k \over 2})({k \over 2}+1-j)  + (m-{k \over 2})^2] + (N-j+m+1/2) +h
}

\noindent
we see that $h^c$ is also strictly positive for
${k \over 2} \leq j < {k\over 2}+1$, and linear independence follows.

In order to see that the states \hilbc\ generate the whole Hilbert space,
let's define as $\cH^{(M)}$ the subspace of states with $L_0$ level $M$. For the states \hilbc,

$$M=N+\sum_{s=0}^a(1/2+s)\varepsilon_s + \sum_{s=1}^m\lambda_ss + \sum_{s=0}^b(1/2+s)\delta_s + \sum_{s=1}^n\mu_ss$$
Let's also define as $\cG^{(M)}$ the subspace of $\cH^{(M)}$ generated by states \hilbc \, such that $M>N$.
We proceed now by induction on $M$ as in \cite{Evans,Goddard}.
For $M=0$, $\cH^{(M)}$ is just formed by the representation $\cD^+_j$ of the zero modes.

Let's assume the induction hypothesis for states with level lower
than $M$. The linear independence argument above has shown that
there are no null states among the descendants \hilbc\ of states
in $\cal{F}$, that is, no states in $\cG^{(M)}$ which are
orthogonal to all states in $\cG^{(M)}$. It follows that
$\cH^{(M)}$ is the direct sum of $\cG^{(M)}$ and its orthogonal
complement.

Consider a state in the orthogonal complement of $\cG^{(M)}$. From the induction hypothesis
that state is annihilated by $L_{n>0}, G_{n>0}, \psi^3_{n>0}$ and $J^3_{n>0}$, that is, it belongs
to $\cal{F}$. It follows that $\cH^{(M)} = \cG^{(M)} \oplus \cF^{(M)}$ where $\cF^{(M)}$ are states in $\cal{F}$
at level $M$, i.e., the states \hilbc\ form a complete basis at each level $M$ and each $J^3_0$ eigenvalue
$m$.

\bigskip
\noindent
{\bf Step 2}: A physical state can be decomposed into the sum of two physical states,
one of them spurious and the other not.

\noindent
{\bf Proof}: A state is called spurious if it is a linear combination of states \hilb\ with $\varepsilon_i \neq 0$ or
$\lambda_i \neq 0$.  Given a physical state expressed
in the basis \hilb, it can be written as a state with no $L_n,G_n$  plus a spurious state. It can be checked
that the action of $L_{n>0},G_{n>0}$ on a spurious state leaves it
spurious provided $L_0= {1 \over 2}$ and $c=15$ \cite{Goddard}. Moreover, the action
of $L_{n>0},G_{n>0}$ on states without $L_{-n},G_{-n}$ won't produce new $L_{-n},G_{-n}$.
Since \hilb\ is a basis, it follows that both the spurious and the nonspurious states satisfy the
physicality conditions \phys\ separately.

Note that since the inner product of a physical and a spurious state vanishes we should only check
unitarity for the state with no $L_{-n},G_{-n}$.

\bigskip

\noindent
{\bf Step 3}: A physical state $|x\rangle$ involving neither $L_{-n}$ nor $G_{-n}$ (when written as \hilb), belongs to $\cal{F}$.

\noindent
{\bf Proof}: Using the decomposition \Tc\ it is clear that $|x\rangle$ is annihilated by
 $L^3_{n>0}$ and $G^3_{n>0}$. A sufficient condition under which $|x\rangle$ will also be annihilated by
 $J^3_{n>0},\psi^3_{n>0}$ is that $m \neq 0$.

The reason is that when $m \neq 0$ the Hilbert space of
$L^3_{-n},G^3_{-n}$ acting on $\cal{F}$ has no null descendants,
because it has a negative highest weight $-{m^2 \over 2k}$, and
null descendants of a $c= {3 \over 2}$ NS supersymmetric Virasoro
algebra only appear for nonnegative highest weights~\cite{Kac}.

Assuming the absence of null Virasoro descendants, the
$L^3_{-n},G^3_{-n}$ Hilbert space and that generated by
$J^3_{-n},\psi^3_{-n}$ will be the same\footnote{This follows by
simply counting the number of states at each level.}. So our
$|x\rangle$ state, which is a $J^3_{-n},\psi^3_{-n}$ descendent,
can be expressed as a $L^3_{-n},G^3_{-n}$ descendent. But being
annihilated by $L^3_{n>0}$ and $G^3_{n>0}$, $|x\rangle$ can only
be the highest weight state, which is also annihilated by
$J^3_{n>0},\psi^3_{n>0}$.

The above argument does not hold for states with $m=0$, and we
treat this case independently. Let's consider the appearance of
physical states with $m=0$ for different values of $j$. Remember
that the mass shell condition is \eqn{mass}{ -{j(j-1) \over k}
+N+h - {1 \over 2}=0 }

\ni
For  $0 <j <1$  there are no states with $m =0$ because $m$ differs from $j$ by integer values.

For  $j =1$, Eq.\mass ~implies that $N=0$ or $N={1 \over 2}$.
In the first case there are no states in ${\cal D}^+_{\textit{j}}$ with $m<j$.
In the second case, the only state is
\eqn{uni}{
\psi^-_{-{1 \over 2}}|j=1,m=1\rangle
}
which is physical and belongs to ${\cal F}$.

For  $j > 1,$ since for $m=0$ we have $N + {1 \over 2} \geq j$, the left hand side of \mass\ is
\eqn{massa}{
-{j(j-1) \over k} +N+h - {1 \over 2} \geq -{j(j-1) \over k} +{k(j-1) \over k} +h
= {(k-j)(j-1) \over k}+h >0
}
because $j < {k \over 2}+1$ implies $(k-j) > {(k-2) \over 2}$, which is positive for $k>2$.
Thus  \mass ~cannot be satisfied either.

It follows that every physical state is annihilated by $J^3_{n>0},\psi^3_{n>0}$ and its
norm is positive by the result of Section 3.

\subsection{Flowed states}

In this case, the states are built by the action of
$\tilde{\psi}^{3,\pm}_{-n}$ and $\tilde{J}^{3,\pm}_{-n}$ on $|\tilde{j},\tilde{t}\rangle$, but
the physicality conditions \phys\ are still imposed with the unflowed operators  $L_n$ and $G_n$.
Since we are considering the $\cD_j^+$ series, we need only consider spectral flow with $w>0$.

\bigskip
\ni
{\bf Step 1}: By the same arguments as before, we know that the states as those in \hilb, but written with
$\tilde{\psi}^3_{-n}, \tilde{J}^3_{-n}, \tilde{L}_{-n},\tilde{G}_{-n}$  form\footnote{Note that the conditions
\primaries\ defining $\cal{F}$ can equivalently be imposed with either flowed or unflowed operators.}
a basis
for states built upon $|\tilde{j},\tilde{t}\rangle$. But replacing in \hilb\ these operators with
\eqn{replace}{
\eqalign{
\tilde{J}_n^3 & = J_n^3  \cr
\tilde{\psi}_n^3 & = \psi_n^3 \cr
\tilde{L}_n & =L_n + wJ^3_n  \cr
\tilde{G}_n & =G_n + w\psi^3_n
}}
for $n<0$, they can be put in one to one correspondence with the states \hilb.
In other words, the states
\hilb, with ${\psi}^3_{-n},{J}^3_{-n}, {L}_{-n},{G}_{-n}$, are a  basis for the Hilbert space built by $\tilde{\psi}^{3,\pm}_{n}, \tilde{j}^{3,\pm}_{n}$
upon $|\tilde{j},\tilde{t}\rangle$.

\bigskip
\ni
{\bf Step 2}: It is the same.

\bigskip
\ni
{\bf Step 3}: The physicality conditions \phys\ for $|x\rangle$ again imply
that $L^3_{n>0}$ and $G^3_{n>0}$ annihilate it, and this in turn means that
also $\tilde{J}^3_{n>0},\tilde{\psi}^3_{n>0}$ annihilate $|x\rangle$ provided \mbox{$m=\tilde{m} +kw/2 \neq 0$.}
Let's see that no states with $m=0$ can appear in the flowed representations by
using the flowed mass shell condition \shellmass,
\eqn{mf}{
-{\jt ( \jt -1) \over k} -wm + {kw^2 \over 4}+ \tilde{N} + h - {1 \over 2}=0
}

\ni
In the following we use that $\jt < {k \over 2} +1 $ implies that
$-{\jt ( \jt -1) \over k}> -{k \over 4} - {1 \over 2}$.

\ni
For $w=1$ the left hand side of \mf\ for $m=0$ is
\eqn{wunoa}{
-{\jt ( \jt -1) \over k} + {k \over 4}+ \tilde{N} + h - {1 \over 2}  >
\tilde{N} + h - {1 \over 2}
}
and using $\mt= -{k \over 2}$ the right hand side of \wunoa\ is
\eqn{wunob}{
 (\Nt -\jt + \mt + 1/2) +\jt - \mt -1 + h = (\Nt -\jt + \mt + 1/2) + \jt + {(k-2) \over 2} + h > 0
}
because $(\Nt -\jt + \mt + 1/2) \geq 0$ and $k>2$, so that \mf\ is not satisfied.

For $w \geq 2$, the left hand side of \mf\ for $m=0$ is
\eqn{wdos}{
-{\jt ( \jt -1) \over k} + {kw^2 \over 4}+ \tilde{N} + h - {1 \over 2} >
{k \over 4}(w^2 -1) -1 + \tilde{N} + h  > {1 \over 2}+ \tilde{N} + h > 0
}
because $k>2$, so that \mf\ is not satisfied either.

Having seen that physical states are annihilated by $\tilde{J}^3_{n>0},\tilde{\psi}^3_{n>0}$, unitarity follows again from
the result of Section 3.

\renewcommand{\baselinestretch}{0.87}

\section{Conclusions}
We have proved the no-ghost theorem for the NS sector of discrete
representations
of fermionic strings
in $AdS_3 \times \cN$. The result is relevant for
a whole family of vacua, such as those
yielding superconformal supersymmetry
in the boundary CFT theory \cite{GKS,GR}.

For flowed fermionic strings in $\cC_j$ representations,
the result of Section 4.2 can be easily generalized,
but then we should prove the
supersymmetric coset unitarity in the $\cC_j$ sector.

The proof of the no-ghost theorem discussed above is a curved
space generalization of the proof for flat space in the old
covariant quantization scheme \cite{Brower,Goddard}. On the other
hand, both for $\cD^{\pm}_j$ and  for $\cC_j$ representations, the
unitarity requirement for the supersymmetric coset can be bypassed
in the proof of the no-ghost theorem by using instead a BRST
quantization scheme. In that case one can rely upon the bosonic
coset unitarity by using the decomposition into $j^a$ and $\psi^a$
currents~\cite{Ari}. In any case, the unitarity of the
supersymmetric coset is relevant by itself, due to the wealth of
models based on it.

\section*{Acknowledgements}
I thank Shmuel Elitzur and Amit Giveon for conversations and the
Instituto de F\'{\i}sica de La Plata (Argentina) for hospitality during
part of this work.

\end{document}